From Ian.Moss@newcastle.ac.ukThu Feb 27 17:58:03 1997
Date: Thu, 27 Feb 97 16:21:05 GMT
From: Ian Moss <Ian.Moss@newcastle.ac.uk>
To: S.M.Goncalves@newcastle.ac.uk
Subject: bh paper

\documentstyle[preprint,aps,epsf]{revtex}
\tighten
\begin{document}
\draft
\preprint{\hfill NCL97--TP1}
\title{BLACK HOLE FORMATION FROM MASSIVE SCALAR FIELDS}
\author{S\'{e}rgio M. C. V. Gon\c{c}alves and Ian G. Moss}
\address{
Department of Physics, University of Newcastle Upon Tyne, NE1 7RU U.K.
}
\date{January 1997}
\maketitle
\begin{abstract}
It is shown that there exists a range of parameters in which
gravitational collapse with a spherically symmetric massive scalar
field can be treated as if it were collapsing dust. This implies a
criterion for the formation of black holes depending on the size and 
mass of the initial field configuration and the mass of the scalar field.
\end{abstract}
\narrowtext
\section{INTRODUCTION}

An important issue in General Relativity is to ascertain whether a
given matter distribution will collapse to form a black hole. To
address one aspect of this question we shall consider a minimally 
coupled massive scalar field with spherical symmetry which starts 
from rest. Massive scalar fields are important in many models of the 
early universe. If any of these fields underwent gravitational 
collapse then the result would be a primordial black hole.

In recent years, gravitational collapse with massless scalar fields and
spherical symmetry has been analysed in some detail
\cite{chop,gold,chop2,brady}.
One result to come out of this is that, when the initial profile of the
scalar field is fixed in terms of a single parameter, there is a
critical parameter value. Beyond the critical value a black hole will
form\cite{chop2}.

Much less is known about the spherically symmetric collapse of massive
scalar fields. In order to obtain an approximate solution of the
Einstein equations we introduce a WKB approximation for the field in
the limit where the mass of the scalar field is large. To leading
order, the solution behaves in the same way as inhomogeneous dust. This
is known already for the linearised Einstein equations, but we find
that the WKB approximation is equally good in the non-linear regime.

The dust solution, like the Oppenheimer-Snyder\cite{oppen} solution,
always collapses to a black hole. By keeping track of the size of the
terms that are higher order in the approximation we obtain a sufficient
condition for the validity of the approximation and therefore for the
gravitational collapse of the massive scalar field. This condition can
be obtained analytically when the initial density has a top-hat form,
and numerically otherwise.

Planck units (in which $G=c=\hbar=1$) are used throughout.

\section{THE MODEL}

The most general spherically symmetric spacetime metric can be written
in the form
\begin{equation}
{\rm d}s^{2}=
-{\rm e}^{2\Psi}{\rm d}\tau^{2}+{\rm e}^{-2\Lambda}{\rm
dr}^{2}+R^{2}{\rm d}\Omega^{2},
\end{equation}
where $\Psi=\Psi(\tau,r)$, $\Lambda=\Lambda (\tau,r)$, $R=R(\tau,r)$
and ${\rm d}\Omega^{2}={\rm d}\theta^{2}+\sin\theta {\rm d}\phi^{2}$
is the metric of a unit 2-sphere. It is convenient, for our purposes,
to rescale the coordinate $\tau$ to the proper time $t$ of a comoving
observer at radial coordinate $r$,
\begin{equation}
t=\int_0^{\tau} {\rm e}^{\Psi (\tau',r)}{\rm d}\tau',
\end{equation}
hence one gets the metric in Gaussian coordinates,
\begin{equation}
{\rm d}s^{2}= -{\rm d}t^{2}+{\rm e}^{-2\Lambda}{\rm d}r^{2}+R^{2}{\rm
d}\Omega^{2}.
\end{equation}

The Einstein tensor components are
\begin{eqnarray}
G_{tt} & = & R^{-2}\left[ -R{\rm
e}^{2\Lambda}(2R'\Lambda'+2R''+R^{-1}R^{\prime2})-2{\dot R}{\dot
\Lambda}R+1+{\dot R}^{2}\right] \\
G_{rt} & = & -2R^{-1}({\dot R'}+R'{\dot \Lambda}) \\
G_{rr} & = & -R^{-2}\left[ {\rm e}^{-2\Lambda}(2{\ddot R}R+{\dot
R}^{2}+1)-R'^{2}\right] \\
G_{\theta\theta} & = & \sin^{-2}\theta G_{\phi\phi} = R[{\dot R}{\dot
\Lambda} + R'\Lambda' e^{2\Lambda}
+R''e^{2\Lambda}-{\ddot R}+{\ddot \Lambda}R-{\dot \Lambda}^{2}R]
\end{eqnarray}
where $\partial_{t}\equiv \;\cdot$, $\partial_{r}\equiv \;'$.

For the matter content, we will take a real minimally coupled scalar
field of mass $\mu$, whose equation of motion is
\begin{equation}
(-\nabla^2+\mu^2)\phi=0.
\end{equation}
In the spherically symmetric metric,
\begin{equation}
\ddot\phi-e^{2\Lambda}\phi''+\mu^2\phi+\left(\dot\Lambda+2R^{-1}\dot
R\right)\dot\phi-e^{2\Lambda}
\left(\Lambda'-2R^{-1}R'\right)\phi'=0\label{sdot}
\end{equation}
The stress-energy tensor of the scalar field has components
\begin{eqnarray}
T_{tt} & = & \frac{1}{2}{\dot \phi}^{2}+\frac{1}{2}{\rm
e}^{2\Lambda}\phi'^{2}+\frac{1}{2}\mu^{2}\phi^{2} \\
T_{rt} & = & {\dot \phi}\phi' \\
T^{ r}_{r} & = & \frac{1}{2}{\dot \phi}^{2}+\frac{1}{2}{\rm
e}^{2\Lambda}\phi'^{2}-\frac{1}{2}\mu^{2}\phi^{2} \\
T_{\theta\theta} & = & \sin^{-2}\theta T_{\phi\phi} =
\case1/2R^{2}({\dot \phi}^{2}-e^{2\Lambda}\phi'^{2}-\mu^{2}\phi^{2})
\end{eqnarray}

It proves convenient to introduce the functions
\begin{eqnarray}
k(t,r) & = & 1-{\rm e}^{2\Lambda}R'^{2}\label{kdef} \\
m(t,r) & = & \case1/2R({\dot R}^{2}+k),\label{mdef}
\end{eqnarray}
so that two Einstein equations can be recast in terms of the first
derivatives of these two functions,
\begin{eqnarray}
{\dot k} & =&  8\pi RR'T^{r}_{t} \label{kdot}\\
{\dot m}  &= & 4\pi R^{2}R'T^{r}_{t}-4\pi R^{2}{\dot
R}T^{r}_{r}\label{mdot}
\end{eqnarray}
These two equations, together with the scalar equation (\ref{sdot}),
form a complete set.  A third Einstein equation provides a constraint
\begin{equation}
m'  =  4\pi R^{2}R'T_{tt}-4\pi R^{2}{\dot R}T_{rt} \label{mprime}
\end{equation}
which is important for relating $m(r,0)$ to the initial data.

The equations have an approximate solution when the compton wavelength
of the field $\mu^{-1}$ is much smaller than the radius $\lambda$ of
the spherical region where the field is non-vanishing,
\begin{equation}
\lambda\mu\gg 1
\end{equation}
The scalar field has a wave-like solution with nearly constant
amplitude,
\begin{equation}
\phi (t,r) = \mu^{-1}\Phi(t,r)\cos(\mu t).
\end{equation}
The stress-energy tensor becomes
\begin{eqnarray}
T_{tt} & = & \case1/2\Phi^2-\case1/2\mu^{-1}\Phi\dot\Phi\sin\;2\mu
t+\case1/4\mu^{-2}(\dot\Phi^2+e^{4\Lambda}\Phi^{\prime2})(1+\cos\;2\mu
t)\\
T_{rt} & = & -\case1/2\mu^{-1}\Phi\Phi'\sin\;2\mu t+
\case1/2\mu^{-2}\dot\Phi\Phi'(1+\cos\;2\mu t) \\
T^{\ r}_{r} & = & -\case1/2\Phi^2\cos\;2\mu
t-\case1/2\mu^{-1}\Phi\dot\Phi\sin\;2\mu
t+\case1/4\mu^{-2}(\dot\Phi^2+e^{4\Lambda}\Phi^{\prime2})(1+\cos\;2\mu
t)\\
T_{\theta\theta} & = & \case1/2R^2\left[-\Phi^2\cos\;2\mu
t-\mu^{-1}\Phi\dot\Phi\sin\;2\mu
t+\case1/2\mu^{-2}(\dot\Phi^2-e^{4\Lambda}\Phi^{\prime2})(1+\cos\;2\mu
t)\right]
\end{eqnarray}

The form of the stress-energy tensor suggests a trigonometric expansion
for the function $\Phi$ of the form
\begin{equation}
\Phi=\Phi_0+\sum_{m=1}^\infty\sum_{n=
2}^{2m}\,\mu^{-m}\left(\Phi^c_{mn}\cos\;n\mu t+\Phi^s_{mn}\sin\;n\mu
t\right)
\end{equation}
with similar expansions for $m$ and $k$.
More precisely, we take
\begin{eqnarray}
\Phi (t,r) & = &  \Phi_{01}(t,r)\cos \mu t + \mu^{-2}\Phi_{22}\cos 2\mu
t + O(\mu^{-3}) \\
m(t,r) & = & m_{0}(t,r) + \mu^{-1}m_{12}(t,r)\sin 2\mu t +
\mu^{-2}m_{22}(t,r)\cos 2\mu t +O(\mu^{-3}) \\
k(t,r) & = & k_{0}(t,r) + \mu^{-2}k_{22}(t,r)\cos 2\mu t + O(\mu^{-3})
\\
R(t,r) & = & R_{0}(t,r) + \mu^{-2}R_{22}(t,r)\cos 2\mu t + O(\mu^{-3})
\end{eqnarray}
Terms which play no role have been dropped.

When the expansions are substituted into eqs. (\ref{mdef}--\ref{mdot})
the leading order terms give
\begin{eqnarray}
\dot m_0&=&0\\
\dot k_0&=&0\\
\dot R^2_0&=&2m_0R_0^{-1}-k_0\\
m_0'&=&2\pi R_0^2R_0'\Phi_0^2
\end{eqnarray}
The first three equations also arise in the Tolman--Bondi solution
representing the collapse of spherically symmetric dust. We deduce that
to leading order the metric will have Tolman--Bondi form.

The next order gives
\begin{eqnarray}
k_{22}(t,r) & = & 2\pi R_{0}(R'_{0})^{-1}(1-k_{0})\Phi_{0}\Phi'_{0} \\
m_{12}(t,r) & = & \pi R_{0}^{2}{\dot R}_{0}\Phi_{0}^{2} \\
R_{22}(t,r) & = & -\case1/2 R_{0}\Phi_{0}^{2},
\end{eqnarray}
We will assume that the WKB approximation holds as long as these
correction terms are less than $1/2$ the leading terms. (The effect of
changing this number will be equivalent to changing the value of
$\mu$). The largest correction comes from the $m_{12}$ term, therefore
the WKB approximation should be valid whilst
\begin{equation}
\mu^{-1}{m_{12}\over m_{0}}\le{1\over 2}
\end{equation}
This inequality defines a region of the $(t,r)$ coordinate plane. The
region is bounded by a curve where the inequality is saturated,
\begin{equation}
{\dot R_0}m'_0 = \mu m_0R'_0.\label{wkb}
\end{equation}
There is no reason to extend the leading order approximation beyond
this curve. Therefore the region of validity of the approximation will
only consist of points in whose causal past the inequality is always
satisfied.

\section{TOLMAN--BONDI METRICS}

We have seen that a sufficiently massive scalar field behaves like
inhomogeneous dust. The corresponding metric can be recovered from the
definitions (\ref{kdef}) and (\ref{mdef}), with $m(t,r)\equiv m(r)$ and
$k(t,r)\equiv k(r)$ (dropping the `0'-indices for simplicity),
\begin{eqnarray}
{\rm d}s^{2} & = & -{\rm d}t^{2}+\frac{R'^{2}(t,r)}{1-k(r)}{\rm
d}r^{2}+R^{2}(t,r){\rm d}\Omega^{2} \\
{\dot R}^{2}(t,r) & = & \frac{2m(r)}{R(t,r)}-k(r) \label{rdt}\\
\Phi^{2}(t,r) & = & \frac{1}{2\pi}\frac{m'}{R^{2}R'},\label{phie}
\end{eqnarray}
The metric has Tolman--Bondi form \cite{tol34,bon48}. This is a well
known class of solutions that can be thought of as a collection of
free--falling spherical shells. Assuming $R'\neq 0$ (then none of the
shells cross), one can analytically integrate equation (\ref{rdt})
parametrically
\begin{eqnarray}
t(\eta,r) & = & t_{0}(r) + \frac{m}{k^{\frac{3}{2}}}(\eta +\sin\eta) \\
R(\eta,r) & = & \frac{2m}{k}\cos^{2}\frac{\eta}{2}
\end{eqnarray}
where $t_{0}(r)$ is an arbitrary function.

We take time-symmetric initial data with
\begin{eqnarray}
R(0,r)&=&r\\
{\dot R}(0,r)&=&0
\end{eqnarray}
These initial conditions fix the radial coordinate $r$ and the
functions $t_0(r)=0$, $k(r)=2m(r)/r$,
\begin{eqnarray}
R(\eta,r) & = & r\cos^{2}{\eta\over 2} \\
t(\eta,r) & = & \sqrt{\frac{r^{3}}{8m}}(\eta + \sin \eta)
\end{eqnarray}
where $\eta$ runs from $0$ to $\pi$. A particular solution is uniquely
specified by $m(r)$ and related to the initial density profile
$T_{tt}(0,r)$,
\begin{equation}
m(r)=4\pi\int_0^r T_{tt}(0,r')r^{\prime2}dr'
\end{equation}
The function $m(r)$ is the total mass within the sphere of radius $r$.
If it approaches a constant value $M$ at large radius then $M$ is the
ADM mass.

\begin{figure}
\begin{center}
\leavevmode
\epsfxsize=20pc
\epsffile{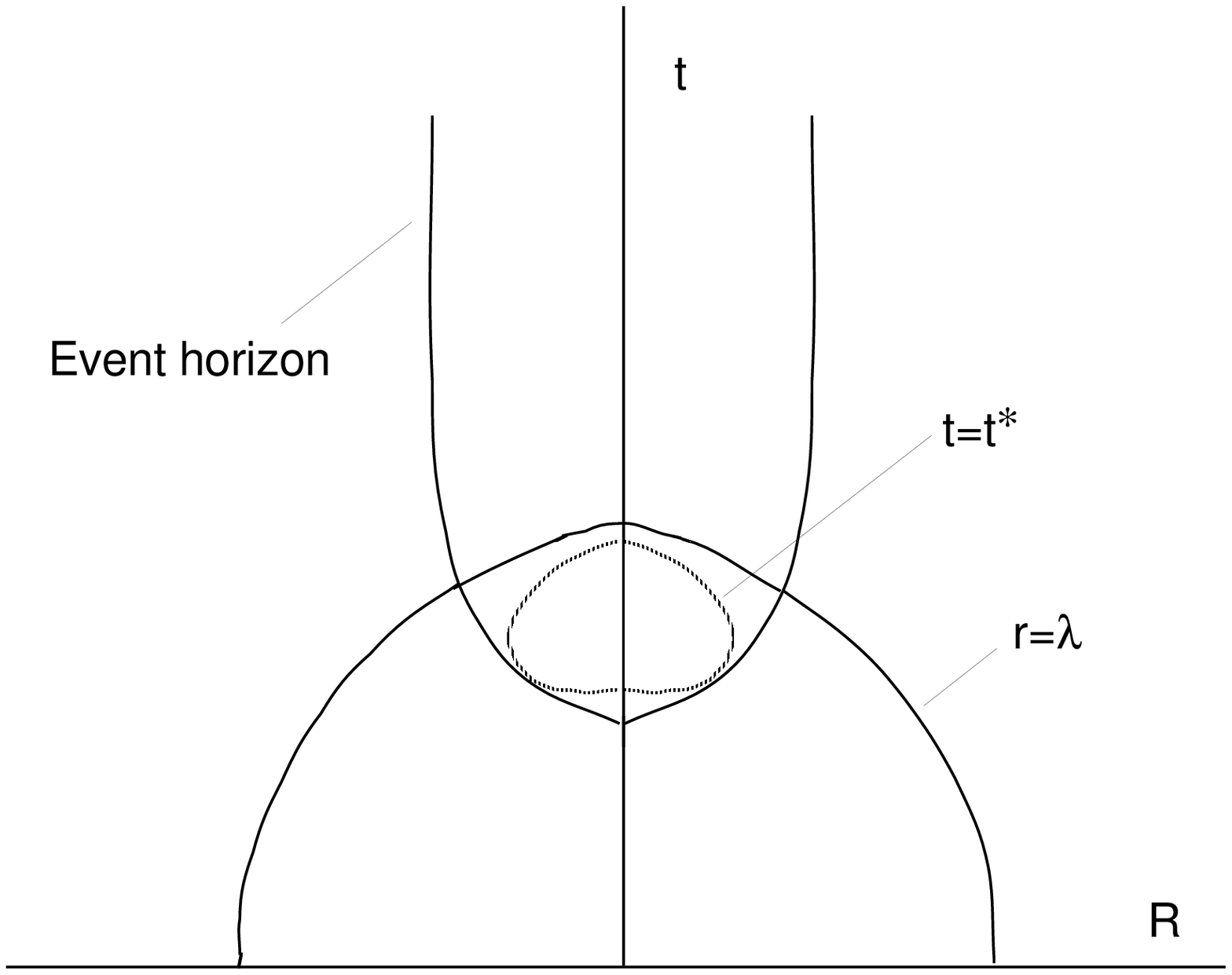}
\end{center}
\caption{A section of the Tolman--Bondi metric with coordinates
$(R,t)$ for an exponential density distribution. The shell $r=\lambda$ 
and the event horizon are shown, as well as the line $t=t_*$ where the WKB
approximation breaks down..\label{fig1}}
\end{figure}

A model for the initial density profile which we make use of later is
\begin{equation}
\rho (r) \equiv T_{tt}(0,r)=\rho_{0}e^{-(r/\lambda)^{3}},\label{exp}
\end{equation}
so that the mass function is simply,
\begin{equation}
m(r)=M(1-e^{-(r/\lambda)^{3}})
\end{equation}
This mass distribution collapses to form a black hole of radius $2M$.
The centre collapses first, and forms a singularity after a proper-time
$\pi/\sqrt{8M}$. This is shown in figure \ref{fig1}

Also shown in figure \ref{fig1} is the event horizon, obtained by
solving the equation for radial null geodesics
\begin{equation}
{{\rm d}\eta\over{\rm d} r} = \left[
\pm\left({r\over2m}-1\right)^{-1/2} R'-{1\over2}\gamma
(\eta+\sin\eta)\right] R^{-1}\label{rng}
\end{equation}
where
\begin{equation}
\gamma(r)={3\over2}-{rm'\over2m}\label{gamma}
\end{equation}
with the boundary condition $R\to2M$ as $r\to\infty$.

Various derivatives of the radial function that are of use are
\begin{eqnarray}
{\dot R} & = &
-\left(\frac{2m}{r}\right)^{\frac{1}{2}}\tan{\eta\over2}\label{rd}\\
R' & = &  \cos^{2}{\eta\over2} +
{1\over2}\gamma\left(\eta+\sin\eta\right)\tan{\eta\over2}\label{rprime}
\end{eqnarray}
The value of $\Phi$ given by eq. (\ref{phie}) will diverge if $R'=0$,
from which we obtain a restriction on the initial mass distribution:
$\gamma\ge0$.

\section{THE OPPENHEIMER--SNYDER LIMIT}

Oppenheimer and Snyder considered the collapse of a star with constant
density, for which
\begin{equation}
m(r)=\cases{\displaystyle
M\left({r\over\lambda}\right)^3&$0<r<\lambda$\cr
M&$r\ge\lambda$}
\end{equation}
In this case the model can be solved explicitly and we can obtain
analytic expressions for the domain of validity of the WKB
approximation. We are especially interested in those cases where the
domain of validity covers the whole region outside the event horizon.

A constant scalar field gives a constant density but the discontinuity
at $r=\lambda$ causes the WKB approximation to break down. It is
possible to change the form of the field close to the edge to make the
field continuous and still have constant density,
\begin{equation}
\Phi=\cases{\Phi_0&$r\le\lambda-\mu^{-1}$\cr
\Phi_1(r)&$\lambda-\mu^{-1}<r<\lambda$\cr
0&$r\ge\lambda$}
\end{equation}
where $\Phi_0=O(1)$ and $\Phi_1'=O(1)$.

\begin{figure}
\begin{center}
\leavevmode
\epsfxsize=20pc
\epsffile{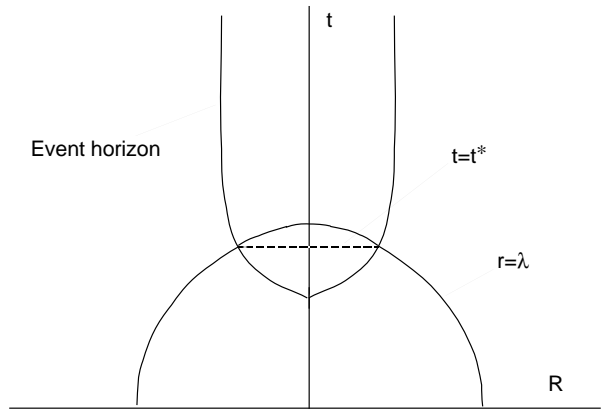}
\end{center}
\caption{A section of the Oppenheimer-Snyder metric with coordinates
$(R,t)$. The edge of the matter distribution $r=\lambda$ and the event
horizon are shown, as well as the line $t=t_*$ where the WKB
approximation breaks down.\label{fig2}}
\end{figure}

The WKB approximation breaks down along the curve given by equation
(\ref{wkb}). Let $\eta_*(r)$ parameterise this curve, then from eqs.
(\ref{rd}) and (\ref{rprime}),
\begin{equation}
\tan{\eta_*\over2}={\mu\over3}\sqrt{\lambda^3\over2M}
\cos^2{\eta_*\over2}.\label{etas}
\end{equation}
Therefore $\eta_*(r)$ is constant for $r<\lambda$. In this case
$t(\eta_*,r)$ is also constant. The Tolman-Bondi solution is exact for
$r>\lambda$, therefore the WKB approximation breaks down along a line
segment from $r=0$ to $r=\lambda$ with fixed time coordinate $t_*$. If
this line segment lies inside the event horizon of the Tolman--Bondi
metric then the WKB approximation must be valid outside of the event
horizon. This is shown in figure \ref{fig2}.

The event horizon is given by a curve $\eta_{EH}(r)$. Outside of the
matter distribution the event horizon lies along $R=2M$, therefore
\begin{equation}
\cos^2{\eta_{EH}(r)\over 2}={2M\over r} \;\;\; r\geq \lambda
\end{equation}
Now, the end of the line where the WKB approximation breaks down will
lie inside the event horizon when
\begin{equation}
\eta_{*}>\eta_{EH}(\lambda)
\end{equation}
which is when
\begin{equation}
\tan{\eta_{*}\over 2}>\sqrt{{\lambda\over2M}-1}
\end{equation}
Now, from equation (\ref{etas}) we have
\begin{equation}
\tan{\eta_{*}\over 2}\left(1+\tan^2{\eta_{*}\over 2}\right)
={\mu\over3}\sqrt{\lambda^3\over2M}.
\end{equation}
The condition that the WKB approximation breaks down inside the event
horizon is therefore
\begin{equation}
\mu M>\frac{3}{2}\sqrt{1-\frac{2M}{\lambda}}.
\end{equation}
If this inequality is satisfied, then the scalar field configuration
with ADM mass $M$ which is initially constant for $r<\lambda$ will
collapse to form a black hole. The arguments presented here say nothing
about what happens when the inequality is not satisfied. The inequality
approaches $\mu M>3/2$ when $\lambda$ is much larger than the initial
Schwarzschild radius.

\section{INHOMOGENEOUS MODELS}

For more complicated density distributions it becomes necessary to test
the validity of the WKB approximation numerically. We will concentrate
here on the exponential density distribution (\ref{exp}) given earlier
and compare this to the results for the top-hat distribution of the
Oppenheimer--Snyder limit.

The aim, as before, is to locate the parameter ranges for which a black
hole forms because the WKB approximation holds good everywhere outside
of the event horizon. The method we use is to find the curve on which
the WKB approximation breaks down and find conditions which place this
curve wholly inside the event horizon.

Differentiating eq. (\ref{wkb}) gives an equation for $\eta_*(r)$,
\begin{equation}
\cos^{2}{\eta_*\over2}
\left[ \gamma -\frac{1}{2}(3+\cot^{2}\frac{\eta_*}{2})\right]
{{\rm d}\eta_*\over{\rm d} r} =
\beta' -{1\over2}\gamma' (\eta_* + \sin \eta_*)
\end{equation}
with $\eta_*(0)=\eta_*$, given by eq. (\ref{etas}) and
\begin{eqnarray}
\gamma(r)&=&{3\over2}-{rm'\over2m}\label{gamma}\\
\beta (r)& =& \left(\frac{2m}{r}\right)^{\frac{1}{2}}\frac{m'}{\mu m}
\end{eqnarray}
This equation can be solved numerically using a simple Runge-Kutta
method. An example is shown in figure \ref{fig1}.

When the shape of initial field configuration is fixed, the model is
parameterised by the scalar field mass $\mu$, the length scale
$\lambda$ and the total mass $M$. The method used here only depends on
the combinations $\lambda\mu$ and $M/\lambda$. The numerical results
for two density distributions are shown in figure \ref{fig3}. Above the
lines, the WKB approximation is valid outside of the event horizon and
a black hole will form. This condition is sufficient but not necessary.
The numerical results for the Openheimer--Snyder limit agree with the
analytical result to six significant figures on the majority of data
points.

\section{CONCLUSIONS}

We have analysed in some detail the regime in which a spherically
symmetric massive scalar field can be treated as if it where dust. The
result is a criterion for the formation of black holes depending on
size and mass scales. The criterion depends to some extent on the form
of the density distribution.

\begin{figure}
\begin{center}
\leavevmode
\epsfxsize=24pc
\epsffile{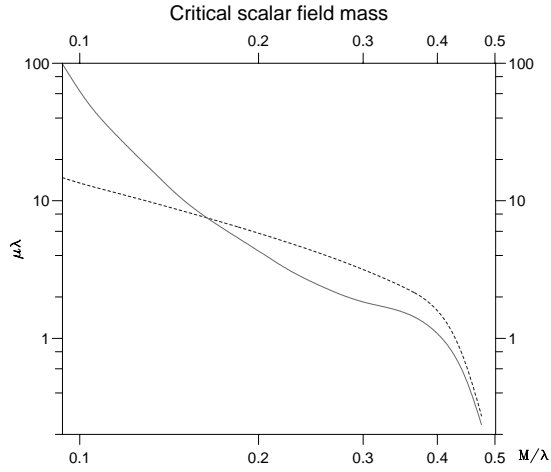}
\end{center}
\caption{Configurations with parameters above the lines will form black
holes. The dotted line denotes the top--hat field profile and the solid
line denotes the exponential field profile discussed in the text. $\mu$
is the scalar field mass, $M$ the ADM mass and $\lambda$ the length
scale.\label{fig3}}
\end{figure}

Outside of the WKB regime the solution of the Einstein equations is a
more difficult problem. The most studied case is the massless limit,
where echoing behaviour was discovered at the point which separates the
cases which form black holes from those which do not. This point lies
on the horizontal axis in figure \ref{fig3}. It is quite possible that
the line which bounds black hole formation continues up from this
point. It certainly must lie under the lines shown in the figure.

In some inflationary models of the early universe, the end of the
inflationary phase is followed by a period dominated by the energy of a
massive scalar field \cite{linde,hawking,albrecht}. It has been
mentioned before that the matter content of the universe behaves like
dust during this period, leading to a situation in which density
fluctuations grow and could form primordial black holes\cite{khlopov}.

The length of time for which the massive scalar field dominates is
limited by the decay of the field into radiation. The Tolman--Bondi 
models can be used to predict how much time is necessary for the 
collapse to a black hole. In an Einstein--de Sitter universe of density
$\rho_c=3H^2/8\pi$, a region of dust with density $\rho=\rho_c(1+\zeta)$ collapses to a singularity in a time $t_c\approx\case1/2\pi H^{-1}\bar\zeta^{-3/2}$, where $\bar\zeta$ is the volume average of 
$\zeta$. If the WKB approximation is valid then a black hole will 
form only if the scalar field has not decayed during the time $t_c$.

The results obtained here for the validity of the WKB approximation
have been limited to asymptotically flat spacetimes and time symmetric
initial data but they can be adapted to remove these limitations for
applications to cosmology. We expect there to be little change to 
the results if the length scale $\lambda$ is smaller than the 
horizon size $1/H$.

We have a simple condition for validity of the WKB approximation in the
Oppenheimer--Snyder limit, $M\mu\gg1$. In the early universe, the mass 
within the scale $\lambda$ is given by $M=\case1/2H^2\lambda^3$. The 
scalar field will behave like dust provided that $H^2\lambda^3\mu\gg1$. 
If we take $H\lambda\sim1$, then the condition becomes 
$\lambda\mu\gg1$, which means that the WKB approximation is valid up 
to the formation of the event horizon if it is valid initially.

\acknowledgments

S.G. is supported by the Programa PRAXIS XXI of the J. N. I. C. T. of
Portugal.


\end{document}